\UseRawInputEncoding 
\documentclass[superscriptaddress,aps,prx,reprint,twocolumn,amsmath,amssymb,showpacs,]{revtex4-2}
\usepackage{graphicx}  % needed for figures
\usepackage{mathptmx}
\usepackage{epstopdf}
\usepackage{dcolumn}   % needed for some tables
\usepackage{bm}        % for math
\usepackage{amssymb}   % for math
\usepackage{amsmath}   % for matrix
\usepackage{amsthm}
\usepackage{chemarrow}
\usepackage{color}
\usepackage{mathrsfs}
\usepackage{float}
\usepackage{bbold}
\usepackage{bbm}
\usepackage[a4paper,colorlinks=true,
linkcolor=blue,citecolor=blue,
pdfauthor={ },
pdftitle={ },
pdfsubject={ },
pdfkeywords={ }]{hyperref}

\theoremstyle{plain}
\newtheorem*{theorem*}{Theorem}

\usepackage{verbatim}

\begin{document}

\title{Multiparameter quantum metrology using strongly interacting spin systems}

\author{Min Jiang}
\affiliation{
Hefei National Laboratory for Physical Sciences at the Microscale and Department of Modern Physics, University of Science and Technology of China, Hefei 230026, China}
\affiliation{
CAS Key Laboratory of Microscale Magnetic Resonance, University of Science and Technology of China, Hefei 230026, China}
\affiliation{
Synergetic Innovation Center of Quantum Information and Quantum Physics, University of Science and Technology of China, Hefei 230026, China}

\author{Yunlan Ji}
\affiliation{
Hefei National Laboratory for Physical Sciences at the Microscale and Department of Modern Physics, University of Science and Technology of China, Hefei 230026, China}
\affiliation{
CAS Key Laboratory of Microscale Magnetic Resonance, University of Science and Technology of China, Hefei 230026, China}
\affiliation{
Synergetic Innovation Center of Quantum Information and Quantum Physics, University of Science and Technology of China, Hefei 230026, China}

\author{Qing Li}
\affiliation{
Hefei National Laboratory for Physical Sciences at the Microscale and Department of Modern Physics, University of Science and Technology of China, Hefei 230026, China}
\affiliation{
CAS Key Laboratory of Microscale Magnetic Resonance, University of Science and Technology of China, Hefei 230026, China}
\affiliation{
Synergetic Innovation Center of Quantum Information and Quantum Physics, University of Science and Technology of China, Hefei 230026, China}

\author{Ran Liu}
\affiliation{
Hefei National Laboratory for Physical Sciences at the Microscale and Department of Modern Physics, University of Science and Technology of China, Hefei 230026, China}
\affiliation{
CAS Key Laboratory of Microscale Magnetic Resonance, University of Science and Technology of China, Hefei 230026, China}
\affiliation{
Synergetic Innovation Center of Quantum Information and Quantum Physics, University of Science and Technology of China, Hefei 230026, China}

\author{Dieter Suter}
\affiliation{
Experimentelle Physik III, Universit$\ddot{\textrm{a}}$t Dortmund, 44221 Dortmund, Germany}

\author{Xinhua Peng}
\email[]{xhpeng@ustc.edu.cn}
\affiliation{
Hefei National Laboratory for Physical Sciences at the Microscale and Department of Modern Physics, University of Science and Technology of China, Hefei 230026, China}
\affiliation{
CAS Key Laboratory of Microscale Magnetic Resonance, University of Science and Technology of China, Hefei 230026, China}
\affiliation{
Synergetic Innovation Center of Quantum Information and Quantum Physics, University of Science and Technology of China, Hefei 230026, China}

\begin{abstract}{
Interacting quantum systems are attracting increasing interest for developing precise metrology.
In particular, the realisation
that quantum-correlated states and the dynamics of interacting systems can lead to entirely new and unexpected phenomena have initiated an intense research effort to explore interaction-based metrology both theoretically and experimentally.
However, the current framework of interaction-based metrology mainly focuses on single-parameter estimations,
a demonstration of multiparameter metrology using interactions as a resource was heretofore lacking.
Here we demonstrate an interaction-based multiparameter metrology with strongly interacting nuclear spins.
We show that the interacting spins become intrinsically sensitive to all components of a multidimensional field when their interactions are significantly larger than their Larmor frequencies.
Using liquid-state molecules containing strongly interacting nuclear spins,
we demonstrate the proof-of-principle estimation of all three components of an unknown magnetic field and inertial rotation.
In contrast to existing approaches,
the present interaction-based multiparameter sensing does not require external reference fields
and opens a path to develop an entirely new class of multiparameter quantum sensors.
}
\end{abstract}

\maketitle

\section{Introduction}

Quantum metrology exploits quantum resources of well-controlled quantum systems to measure small signals with high sensitivity, and
has great potential for both fundamental science
and concrete applications~\cite{giovannetti2011advances,degen2017quantum,pezze2018quantum,braun2018quantum,aasi2013enhanced, budker2007optical, safronova2018search,lang2015dynamical}.
%technologies ranging from time and frequency estimation~\cite{hinkley2013atomic}, gravitational wave detection~\cite{aasi2013enhanced}, and magnetometry~\cite{kominis2003subfemtotesla, budker2007optical} to tests of fundamental physics~\cite{demille2017probing, safronova2018search}.
%In the most common scheme~\cite{degen2017quantum}, a collection of probe systems evolve under the action of a Hamiltonian containing an unknown parameter, and are measured for estimating the parameter.
%Monitoring the precession of an ensemble of spin systems can measure energy shift at the standard quantum limit.
Interacting quantum systems have been recently recognized as an unprecedented playground to develop precise metrology~\cite{alvarez2015localization, lucchesi2019many,kong2020measurement,peyronel2012quantum,dooley2018robust,nolan2017optimal,zhou2020quantum,frerot2018quantum},
where the interactions between probe systems can be a valuable quantum resource to improve the precision and other metrological performance,
such as the achievement of metrological sensitivity beyond the Heisenberg limit~\cite{roy2008exponentially,napolitano2011interaction,boixo2007generalized,zwierz2010general}, the robustness against thermal noise~\cite{dooley2018robust},
criticality-enhanced metrology~\cite{frerot2018quantum, liu2021experimental,chu2021dynamic,rams2018limits}, and the evasion of rapid thermalization~\cite{rovny2018observation,kominis2003subfemtotesla,kong2020measurement}.
%{\color{red}where interactions among probes can correlate the probes and modify their energy levels.}
Moreover, the use of interactions that correlate probe systems can relax the experimental challenge of preparing entangled probe states~\cite{nolan2017optimal,boixo2008quantum}.

However,
the current framework of interaction-based quantum metrology mainly focuses on single-parameter estimations~\cite{boixo2007generalized,dooley2018robust, napolitano2011interaction, nolan2017optimal,roy2008exponentially, zhou2020quantum,rovny2018observation,frerot2018quantum,liu2021experimental,zwierz2010general},
neglecting many possible applications under realistic conditions.
%There is a strong motivation to further expend to the multiparameter estimation,
Indeed, numerous metrological applications are intrinsically multiparameter,
such as measuring electric, magnetic, or gravitational fields.
As a consequence, the field of multiparameter metrology has been intensively studied both theoretically and experimentally~\cite{szczykulska2016multi, vidrighin2014joint,hou2021zero, roccia2018multiparameter,hou2020minimal,seltzer2004unshielded, patton2014all, thiele2018self,li2017dual,liu2017control}.
The potential combination of interaction-based metrology and multiparameter estimation would open up exciting possibilities for developing precise metrology in applied and fundamental physics,
such as quantum imaging~\cite{spagnolo2012quantum,genovese2016real}, sensor networks~\cite{komar2014quantum},
navigation~\cite{donley2010nuclear, walker1997spin,kornack2005nuclear}, the detection of vector fields~\cite{seltzer2004unshielded, patton2014all, hurwitz1960proton, thiele2018self, li2017dual},
as well as the searches for spin-gravity coupling and ultralight dark matter axions~\cite{wu2018nuclear,garcon2019constraints,jiang2021search}.
Despite these appealing features, a demonstration of multiparameter metrology using the resource of interactions was heretofore lacking.

\begin{figure*}[t]  %htb
	\makeatletter
	\def\@captype{figure}
	\makeatother
\centering
	\includegraphics[scale=0.65]{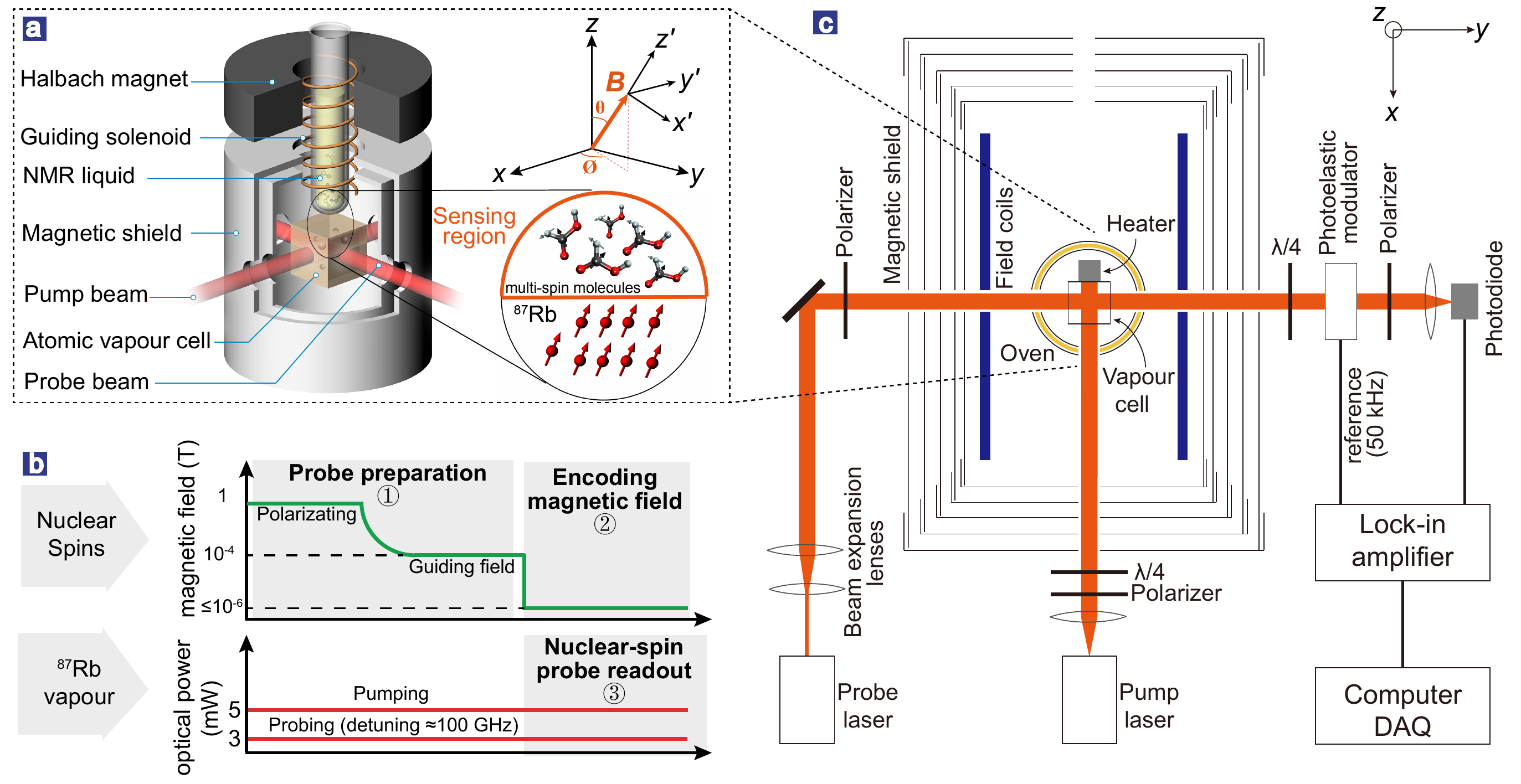} %0.38
	\caption{\textbf{Experimental schematic of interaction-based multiparameter metrology}. \textbf{a}, Diagram of the experimental set-up. The polar and azimuthal angles ($\theta$, $\phi$) parameterize the orientation of the unknown magnetic field $\mathbf{B}(\theta, \phi)$ in the laboratory frame. The information of $\mathbf{B}(\theta, \phi)$ is encoded into the nuclear magnetization of molecules comprising of strongly interacting nuclear spins, which are read out by an atomic vapour sensor. \textbf{b}, Basic procedure of multiparameter metrology, including probe state preparation, encoding unknown magnetic field, and nuclear-spin probe readout. After being polarized in a Halbach magnet, the nuclear spins in liquid-state molecules contained in a 5-mm tube is pneumatically shuttled to a sensing and readout region (at a distance of 1~mm above the atomic vapour); During the pneumatic transfer, the guiding solenoid and three pairs of Helmholtz coils (not shown, at the bottom of the guiding solenoid) are used to generate a guiding field ($\approx 10^{-4}$~T) along an axis (${x}$, ${y}$ or ${z}$) for initializing the spin state; The nuclear spin magnetization of the liquid, which is read out with a $^{87}$Rb vapour sensor (\textbf{c}, see Appendix~A1 for more details), evolves under the unknown magnetic field; Then, the obtained strongly-interacting nuclear-spin spectrum is used to determine the magnitude and orientation of the magnetic field.}
	\label{fig1}
\end{figure*}

Here, we demonstrate an approach to interaction-based multiparameter estimation with the use of strongly interacting nuclear spins in molecules.
%Interactions among nuclear spins correlate their states and energies, leading to enhanced metrology.
We find that,
as the interactions among nuclear spins become significantly larger than their Larmor frequencies,
a different regime emerges where the strongly interacting spins can be simultaneously sensitive to all components of a multidimensional field, such as three-dimensional magnetic field and inertial rotation.
Moreover, we observe that strong interactions between nuclear spins can increase their quantum coherence times as long as several seconds, leading to enhanced measurement precision.
%Moreover, quantum coherence times of the order of a few seconds for strongly interacting spins are observed in our experiment that is an important step towards the realization of high measurement precision.
%all three components of a magnetic field or inertial rotation.
We experimentally illustrate the power of interaction-based multiparameter estimation by applying this approach to simultaneously estimating all three components of a magnetic field.
%and by demonstrating the feasibility of realizing a picotesla-level precision.
%a magnetic field vector with a precision of $10^{-11}$~T and $\sim 0.02$~rad orientation resolution.
%Moreover,
%magnetic pseudo-fields originating from inertial rotation can also be precisely measured by nuclear spins used as a vector gyroscope.
Our work also provides a recipe for precisely measuring magnetic pseudo-fields originating from inertial rotation and constituting a new zero-field vector gyroscope.
The present interaction-based metrology should be generic and promises to develop an entirely new class of multiparameter quantum sensors using a broad range of strongly interacting quantum systems,
and will impact diverse metrological applications.

We would like to emphasize the difference of multiparameter metrology between this work and others.
Although a variety of multiparameter quantum sensors have demonstrated the capability of measuring magnetic fields and inertial rotations~\cite{seltzer2004unshielded, patton2014all, hurwitz1960proton, thiele2018self, kornack2005nuclear, li2017dual,hou2021super,liu2017control},
they all make use of non-interacting probe systems, such as electron and nuclear spins.
For example,
all nuclear magnetic resonance ($\textrm{NMR}$) magnetometers tested so far use only a single spin species (for example, proton rich substances or noble gas $^3$He) that yields a single resonance line at the precession frequency~\cite{legchenko2002nuclear, gross2016dynamic, waters1955measurement}.
Therefore, these magnetometers measure only the magnitude of the magnetic field,
providing no information on its orientation.
In conventional methods, external reference fields are additionally required to access the orientation information~\cite{seltzer2004unshielded,patton2014all,thiele2018self,li2017dual}.
This unavoidably produces significant static and radio frequency fields contamination,
which are classical noise and would be a bottleneck for reaching the fundamental precision beyond the standard quantum limit.
%This reinforces the need to develop xxx.
Unlike previous works,
our work uses a different scheme in which strongly interacting spins are intrinsically sensitive to multidimensional fields.
Thus, our work can avoid the classical noise caused by previously-required reference fields, thus opening a feasible route towards realizing multiparameter quantum sensors with quantum noise-limited sensitivity.
%enabling a new approach to realizing multiparameter sensors.

\section{Interaction-based multiparameter estimation scheme}
Nuclear and electron spin systems are progressively becoming a paradigm for precision metrology.
Here nuclear-spin systems are used as a platform to develop multiparameter metrology.
We focus on the estimation of all three components of an unknown magnetic field.
In contrast to existing works based on non-interacting spins~\cite{seltzer2004unshielded, patton2014all, hurwitz1960proton, thiele2018self, kornack2005nuclear, li2017dual},
we employ a collection of molecules,
where each molecule contains at least two different types of nuclear spins that interact with each other by an exchange interaction (a Heisenberg interaction).
%we perform identical, independent experiments in parallel on each $N$-spin cluster,
%We consider the nuclear spins in liquid-state molecules.
When such nuclear spins are imposed in a magnetic field $\mathbf{B}(\theta, \phi)$,
they can be described by the Hamiltonian,
\begin{equation}
\begin{cases}
\mathcal{H} (\theta, \phi, B)  = \mathcal{H}_{\textrm{spins}} + \mathcal{H}_{\textrm{int}},\\
\mathcal{H}_{\textrm{spins}}=-\sum_j 2\pi \gamma_j \mathbf{I}_j \cdot \mathbf{B}(\theta,\phi),\\
\mathcal{H}_{\textrm{int}}=\sum_{j>i}^n 2\pi {{ J_{ij}}}  \mathbf{I}_i \cdot \mathbf{I}_j,
\end{cases}
\label{Hamiltonian}
\end{equation}
where $\gamma_j$ denotes the gyromagnetic ratio of the $j$th spin (here we set $\hbar=1$),
$\mathbf{I}_j = (\hat{I}_{jx}, \hat{I}_{jy}, \hat{I}_{jz})$ is the spin angular momentum operator of the $j$th spin,
$\mathbf{I}_i \cdot \mathbf{I}_j$ is called a Heisenberg interaction,
and ${J_{ij}}$ is the strength of the Heisenberg interaction between the $i$th and $j$th spins.
As shown in Fig.~\ref{fig1}\textbf{a},
the unknown magnetic field
$\mathbf{B}(\theta,\phi)=(B \sin \theta \cos \phi, B \sin \theta \sin \phi, B \cos \theta)$ is parameterized with the polar angle $\theta$, azimuthal angle $\phi$ and magnitude $B$,
which are unknown parameters to be estimated.
%{\color{red}Specifically, we use $^{13}$CH$_n$-type molecules as the probe system,}
%where each $^1$H spin couples to a $^{13}$C spin with the same scalar spin-spin coupling strength $J$.
%{\color{red}In the presence of a large applied magnetic field,
%which truncates other spin interactions by defining a preferred spatial direction.}
%{\color{red}The described Hamiltonian~(\ref{Hamiltonian}) is ubiquitous in a broad range of spin systems,such as liquid-state and solid-state molecules, quantum dots, and donor-atom nuclear spins.}

%The good quantum numbers are the total spin angular momentum $f$ and the magnetic quantum number $m_f$~[\onlinecite{ledbetter2011,Appelt2010}]
%(see $\textrm{Methods}$ and examples in Fig.~\ref{fig2}).
%{\color{red}We show below that the strongly coupled spin cluster can be intrinsically sensitive to the magnetic field orientation.}
%As we show below, monitoring the spin magnetization of strongly interacting nuclear spins is crucial to estimate all components of the unknown magnetic field.

%The operation procedure of measuring nuclear spin magnetization is shown in Fig.~\ref{fig1}\textbf{b}.
Our interaction-based multiparameter estimation includes three stages: probe preparation, encoding with the strongly interacting spins, and probe readout,
as shown in Fig.~\ref{fig1}\textbf{b}.
The initial probe state of nuclear spins is prepared as $\rho_0$, which is described in Appendix C1 for details.
In the encoding stage,
the key ingredient is the use of the quantum dynamics of the strongly interacting spins to encode the unknown parameters,
\begin{equation}
    \rho(\theta, \phi, B, t)=e^{-i \mathcal{H} (\theta, \phi, B)t}\rho_0 e^{i \mathcal{H} (\theta, \phi, B)t},
    \label{erho}
\end{equation}
where $\rho(\theta, \phi, B, t)$ is the time-dependent density matrix of nuclear spins.
Then, an atomic vapour sensor as a detector
continuously measures the time-evolution of the nuclear magnetization \mbox{$\textrm{Tr} [\rho(\theta, \phi, B, t) \hat{\mathbf{O}}_z]$}~\cite{jiang2018experimental,jiang2019magnetic, tayler2017invited, jiang2020interference}.
Here, $\hat{\mathbf{O}}_z=\sum_j \gamma_j {\hat{I}}_{jz}$ represents the observable operator of $z$-magnetization.
The observation values can be Fourier transformed into a spectrum.
%We can build the relationship between the spectrum of the strongly-interacting nuclear spins and $\mathbf{B}(\theta,\phi)$ in the Hamiltonian $\mathcal{H}(\theta,\phi,B)$.
%the reconstruction of the Hamiltonian $\mathcal{H}(\theta,\phi)$ of the sensor molecules.
Consequently, the information of $\mathbf{B}(\theta,\phi)$ in the Hamiltonian $\mathcal{H}(\theta,\phi,B)$ can be encoded into the spectrum of the interacting spins,
where each spectrally separating line corresponds to the transition between relevant eigenstates $| \Psi_i \rangle$ and $| \Psi_j \rangle$ of $\mathcal{H}(\theta,\phi,B)$
and its amplitude is proportional to a product of $\langle \Psi_i |\rho_0 |\Psi_j \rangle \langle \Psi_j |\hat{\mathbf{O}}_z | \Psi_i \rangle$ [see Eqs.~(\ref{signal}) and (\ref{amp}) in Appendix].
%Here, the initial nuclear spin state is denoted as  $\rho_0$ and the observable operator of nuclear magnetization $\hat{\mathbf{O}}_l=\sum_j \gamma_j {\hat{I}}_{j,l}$ ($l=x,y,z$).

\begin{figure*}[t]  %htb
	\makeatletter
	\def\@captype{figure}
	\makeatother
\centering
	\includegraphics[scale=1.47]{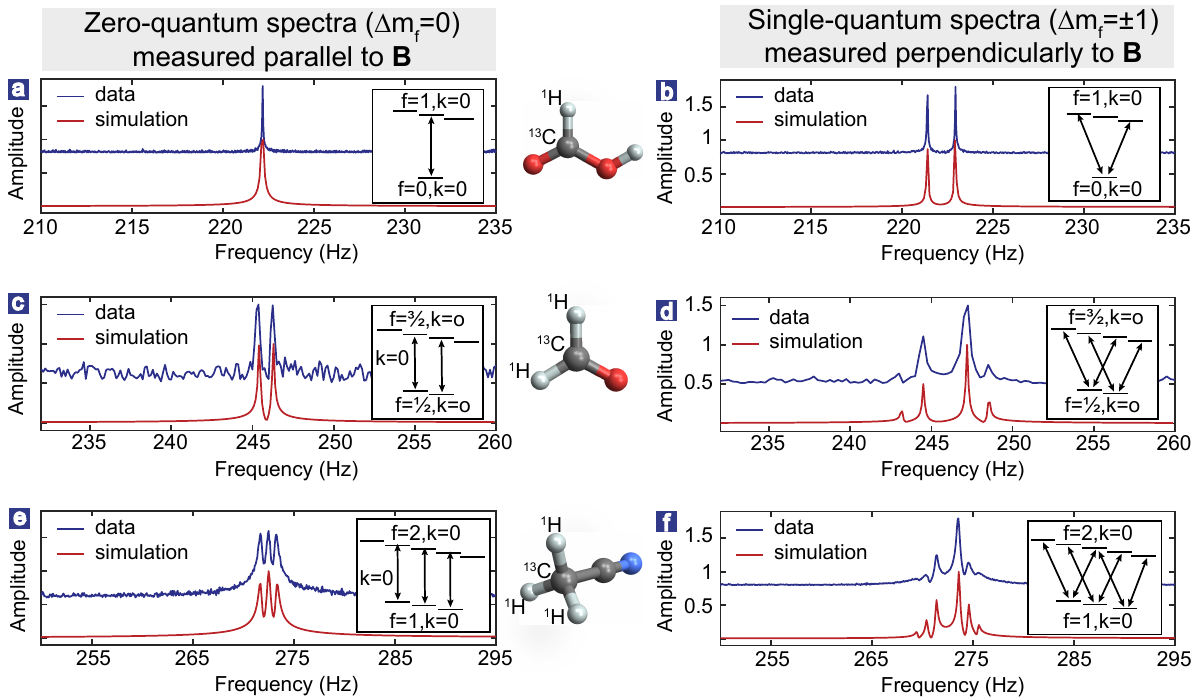}
	\caption{\textbf{Spectra of $^{13}$CH$_n$ molecules measured at various orientations of the magnetic field}. The initial spin state is prepared by the guiding field along the $z$ axis. (\textbf{a}, \textbf{b}) Formic acid (H$^{13}$COOH, where the acidic proton is negligible due to rapid exchange), (\textbf{c}, \textbf{d}) formaldehyde ($^{13}$CH$_2$O), (\textbf{e}, \textbf{f}) acetonitrile ($^{13}$CH$_3$CN)  are detected parallel ($\theta=0$) and perpendicularly ($\theta=\frac{\pi}{2},\phi=0$) to the applied magnetic field. The splitting patterns match well the transitions shown in the inset (see Appendix B1 for details). The amplitudes of the experimental spectra (blue lines) are asymmetric~\cite{jiang2020interference} and are in good agreement with those of the simulated spectra (red lines).}
	\label{fig2}
\end{figure*}

To show the explicit dependence on $\theta$ and $\phi$ of the spectral amplitudes,
we introduce a convenient orthonormal coordinate system,
where
$\mathbf{z'}=\mathbf{B}/B$,
$\mathbf{x'}=\frac{\partial \mathbf{z'}}{\partial \theta}$,
and
\mbox{$\mathbf{y'}=(\frac{1}{\sin \theta})\frac{\partial \mathbf{z'}}{\partial \phi}$}~(see Appendix~C2 for details), as shown in Fig.~\ref{fig1}\textbf{a}.
Consequently,
\mbox{$\langle \Psi_i |\hat{\mathbf{O}}| \Psi_j \rangle=\mathcal{P}(\theta,\phi) \langle \Psi_i |\hat{\mathbf{O}'}| \Psi_j \rangle$},
where $\mathcal{P}(\theta,\phi)$, depending on $\theta$ and $\phi$, is a $3\times3$ transformation matrix [see Eq.~(\ref{P})]
and $\hat{\mathbf{O}}'_{l'}=\sum_j \gamma_j \hat{I}_{jl'}$ ($l'=x',y',z'$).
%The transformation is the same for the initial state $\rho_0$.
Importantly, we note that
the transition amplitudes in the new defined coordinates $\langle \Psi_i |\hat{\mathbf{O}}'| \Psi_j \rangle$
are independent of the magnetic field orientation ($\theta,\phi$) [see Eq.~(\ref{CH})],
where $\langle \Psi_i |\hat{\mathbf{O}}'_{z'}| \Psi_j \rangle$ and $\langle \Psi_i |\hat{\mathbf{O}}'_{x',y'}| \Psi_j \rangle$
respectively correspond to the observation parallel or perpendicular to the magnetic field and hence the zero- or single-quantum resonance line.
Thus, the observed amplitudes of zero- and single-quantum lines therefore can be used to estimate the multiple parameters of $\theta$, $\phi$ and $B$ [see Eq.~(\ref{R}) in Appendix],
and in turn the magnetic field vector $\mathbf{B}(\theta,\phi)$.

We now explain why the strong interaction is a valuable resource for multiparameter estimation.
The situation is considered where the Heisenberg interactions are much stronger than those Larmor frequencies ($|J |\gg |\gamma_j B|$);
in this situation, the nuclear spins in each molecule strongly interact with each other.
Strong interactions between spins can modify the energy level structure of the system,
enabling the direct detection of some forbidden transitions that are pivotal for our multiparameter estimation.
%, in particular, heteronuclear zero-quantum transitions.
To show in more detail, we focus on two-spin case.
The eigenstates of the two-spin system in an arbitrary magnetic field are \mbox{$|\Psi_1 \rangle=\mid\uparrow\uparrow \rangle$},
\mbox{$|\Psi_2 \rangle=\cos \xi\mid\uparrow\downarrow \rangle+ \sin \xi \mid\downarrow\uparrow \rangle$},
\mbox{$|\Psi_3 \rangle=\cos \xi \mid\uparrow\downarrow \rangle$ $- \sin \xi \mid\downarrow\uparrow \rangle$},
$|\Psi_4 \rangle=\mid\downarrow\downarrow \rangle$~[\onlinecite{bene1980nuclear}],
where $\mid \uparrow \rangle$ and $\mid \downarrow \rangle$ are the eigenstates of spin operators $\hat{I}_{jz'}$
along the quantization axis $z'$ (defined by the direction of magnetic field, see Fig.~\ref{fig1}\textbf{a}) and \mbox{$\tan(2\xi)=\frac{ J}{(\gamma_1-\gamma_2)B}$}.
%The scalar $^{13}$C-$^1$H coupling is $J\approx 222.2$~Hz for formic acid.
The transition between $|\Psi_2 \rangle$ and $|\Psi_3 \rangle$ corresponds to a zero-quantum transition,
and the transition probability is $p=|\langle \Psi_2 |\hat{\mathbf{O}}_{z'}|\Psi_3\rangle |=1/2 |(\gamma_1-\gamma_2)\sin (2\xi)|$.
%We now consider the probabilities $p$ for weakly coupled and strongly coupled spin clusters.
We immediately see that the strong interacting regime \mbox{($|J| \gg |\gamma_1 B|,|\gamma_2 B|$)} is of particular interest,
since $|\Psi_2 \rangle$ and $|\Psi_3 \rangle$ are entangled states between two spins and their transition probability \mbox{$p\rightarrow |\gamma_1-\gamma_2|/2$}.
For non-interacting ($J=0$) and weakly interacting ($|J| \ll |\gamma_1 B|,|\gamma_2 B|$) cases,
$p\rightarrow 0$, so the zero-quantum line should never be detected directly.
In contrast, strongly interacting spins can produce oscillating magnetization along the magnetic-field direction ($z'$).
In this situation, zero-quantum spectral lines are detectable and moreover they are very narrow lines due to long-lived coherence mechanism~\cite{sarkar2010long}.
Based on these unforbidden transitions in strongly interacting spin systems that encode unknown parameters,
\mbox{multiparameter estimation becomes possible}.

%Our analysis presents the he strong interacting regime and consider the its implications for estimating all three components of an unknown magnetic field.

%It is therefore difficult to extend this approach into the high-field region
%although nuclear spin clusters have already been measured in conventional high-field NMR.
%{\color{red}add H2O proton magnetometer, >0.1T, weakly}
%In particular, if there are no interactions between nuclear spins ($J=0$),

\section{Spectra of strongly interacting spin systems}

%Our estimation process can be divided into three stages: probe preparation, interaction with the strongly interacting nuclear spins, and probe readout,
%as shown in Fig.~\ref{fig1}\textbf{b}.

It is important to experimentally measure the transition lines of the strongly interacting nuclear spins under the measured magnetic field.
We consider a star topology with one central nuclear spin, and $n$-1 equivalent spins in liquid-state molecules,
such as $^{13}$CH$_n$ molecules,
as the testbed.
A liquid of such molecules is contained in a glass tube with a 100-$\mu$L volume
and polarized with a permanent magnet ($\approx 1.3$~T).
Then the liquid is pneumatically shuttled to a sensing region $\sim 1$~mm above the atomic vapour cell (shown in the inset of Fig.~\ref{fig1}\textbf{a}).
During the shuttling, a guiding field ($\approx 10^{-4}$~T) is applied on the spins,
allowing adiabatic transfer of the polarized nuclear spins into the sensing region and preserving the spin polarization~\cite{jiang2018experimental, jiang2019magnetic, tayler2017invited,jiang2020interference}.
The nuclear-spin coherence times ($\tau_{\textrm{coh}}$) of the spin clusters are typically a few $\textrm{seconds}$ (see Appendix~A2).
As a result, the spin polarization is initialized through controlling the direction of the guiding field [see Eq.~(\ref{rho0}) in Appendix].
After the guiding field is turned off within 10~$\mu$s,
subsequent evolution of the spin state produces oscillating magnetization under the magnetic field to be estimated [see Eq.~(\ref{signal})].
The magnetization encodes the information about the magnetic field
and is read out by an atomic vapour sensor (Fig.~\ref{fig1}\textbf{c}, see Appendix for details).

\begin{figure*}[t]  %htb
	\makeatletter
	\def\@captype{figure}
	\makeatother
\centering
	\includegraphics[scale=1]{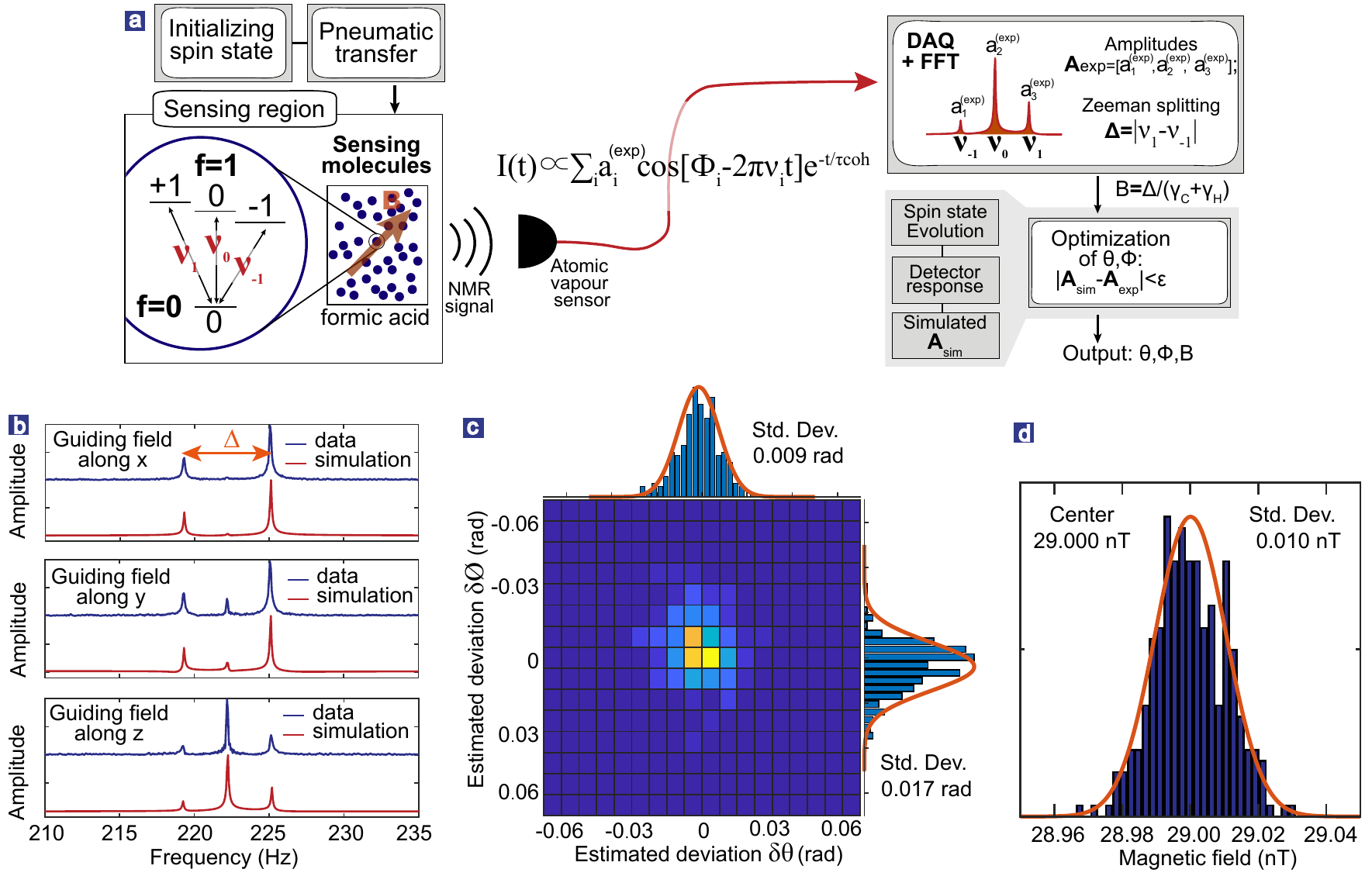}
	\caption{\textbf{Magnetic field sensing with formic acid molecules}. \textbf{a}, The procedure of determining a magnetic field vector. \textbf{b}, As one example of reconstructing magnetic field vector, three initial spin states for formic-acid sensor are independently prepared by applying a guiding field along ${x}$, ${y}$ and ${z}$, in turn. Then the guiding field is immediately turned off. The corresponding time-domain signals under an estimated $B(\theta,\phi)$ are transformed into Fourier spectra (blue lines). The magnetic-field-vector reconstruction yields $\theta \approx 1.289$~rad, $\phi \approx0.047$~rad and $B\approx1.0788\cdot 10^{-7}$~T. The theoretical simulations are shown in red lines. \textbf{c}, Estimated precision of $\theta$, $\phi$ when spectral amplitude suffers from Gaussian noise $\sim \mathcal{N}(0,\sigma ^2)$. Here the variance $\sigma=0.01$ for simulation (corresponding to SNR=100) and the uncertainty $\sigma_{\theta}\approx0.009$ rad and $\sigma_{\phi}\approx0.017$~rad. \textbf{d}, Histograms of magnetic field strength measurements. The measurement of a constant applied $2.9\cdot 10^{-8}~\textrm{T}$ field is repeated 300 times (30~s for each measurement). The precision of measuring the magnetic field magnitude is given by the corresponding standard deviations $10^{-11}$~T.}
	\label{fig4}
\end{figure*}

We first measure different quantum transitions of $^{13}$CH$_n$ molecules.
When one measures the spins along an axis parallel to the magnetic field $\mathbf{B}$ (i.e., $\theta=0$),
the selection rules for allowed transitions are (referred as zero-quantum transitions): $\Delta f=0,\pm 1$, $ \Delta m_f=0$.
For example, formic acid yields a single resonance line at $222.2$~Hz (Fig.~\ref{fig2}\textbf{a}).
On the other hand, if one measures the spins along an axis perpendicular to the magnetic field $\mathbf{B}$ (i.e., $\theta=\pi/2$),
the selection rules for allowed transitions are (referred as single-quantum transitions): $\Delta f=0,\pm 1$, $ \Delta m_f=\pm 1$.
In this case, formic acid yields two resonance lines centered at 222.2~Hz (Fig.~\ref{fig2}\textbf{b}) split by $(\gamma_h+\gamma_c)B$.
Here, gyromagnetic ratios $\gamma_c\approx 10.7077$~MHz/T for carbon spins and $\gamma_h \approx 42.5775$~MHz/T for proton spins.
Spectra of other $^{13}$CH$_n$ molecules show similar patterns, as shown in Fig.~\ref{fig2}\textbf{c}-\textbf{f}.
The detailed splitting patterns are analysed in Eqs.~(\ref{splitting-1})-(\ref{splitting-4}) in Appendix~B1.
Additionally, zero- and single-quantum transitions can be simultaneously allowed
when one measures the spins along an axis neither parallel nor perpendicular to the magnetic field.
This implies that the information of magnetic-field orientation can be extracted from the observed spectral lines of $^{13}$CH$_n$ molecules.
Similar phenomena have been explored in the context of optical spectroscopy of multi-electron atoms~\cite{zeeman1897effect, condon1935theory} but,
to our knowledge, our work represents the \mbox{first experimental demonstration in NMR spectroscopy}.

\begin{figure*}[htp]  %htb
	\makeatletter
	\def\@captype{figure}
	\makeatother
\centering
	\includegraphics[scale=0.727]{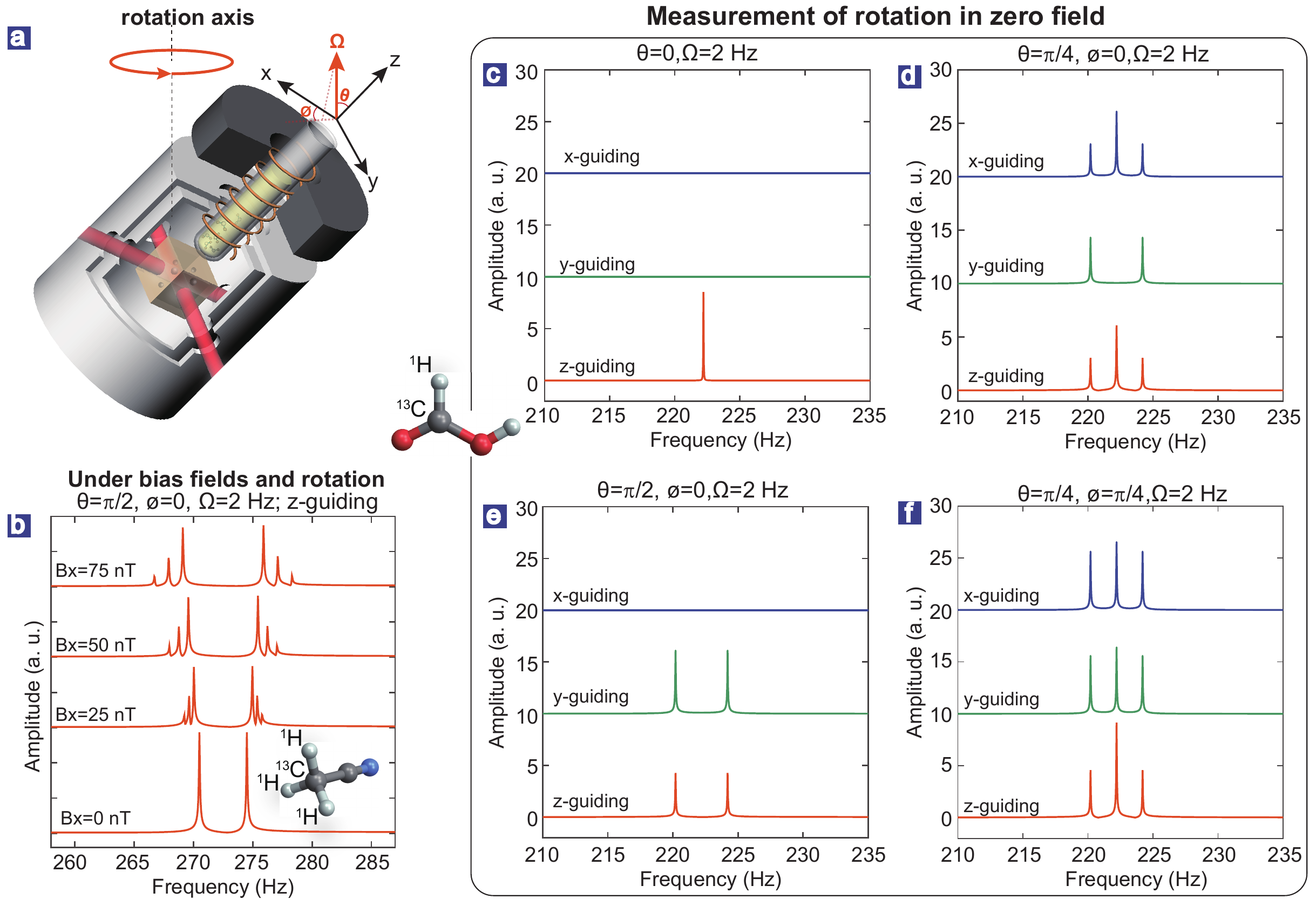}
	\caption{\textbf{Scheme for a vectorial quantum spin gyroscope based on strongly interacting nuclear spins}. \textbf{a}, The proposed vectorial quantum spin gyroscope is the same as the set-up of Fig.~\ref{fig1}\textbf{a}, which operates in zero magnetic field and experiences an inertial rotation $\bm{\Omega}$. The rotation vector $\bm{\Omega}$ can be obtained from the rotationally induced spectra. \textbf{b}, Simulated rotation-induced spectra of acetonitrile ($^{13}$CH$_3$CN) in an inertial rotation (with the rotation frequency of 2~Hz) and various conventional magnetic field strengths along the $x$ axis. \textbf{c,d,e,f}, Measurement of inertial rotations using formic acid (H$^{13}$COOH) for different orientations of the rotation axis in zero magnetic field. The rotation frequency $\Omega=2$~Hz, and the rotation orientations are shown in the corresponding inset.}
\label{rotation}
\end{figure*}

\section{Demonstration of multiparameter metrology}

%\subsection{Magnetometer}
As a proof-of-principle demonstration,
we use formic acid to estimate all three components of a magnetic field.
As shown in Fig.~\ref{fig4}\textbf{a},
in one single scan measurement,
the amplitudes of three spectral lines at frequency $\nu_{0}$~for zero-quantum and at frequency $\nu_{\pm 1}$ for single-quantum are
simultaneously recorded by the normalized form of $\mathbf{A}_{\textrm{exp}}=[\textrm{a}_1^{(\textrm{exp})}(\nu_{-1}), \textrm{a}_2^{(\textrm{exp})}(\nu_0),\textrm{a}_3^{(\textrm{exp})}(\nu_{1})]$.
The magnetic field induces the Zeeman splitting $\Delta=|\nu_1-\nu_{-1}|$ of the doublet and, accordingly,
the magnitude of the measured field can be determined in advance through $\Delta=(\gamma_{c}+\gamma_{h})B$ [the splittings $\Delta$ for arbitrary $^{13}$CH$_n$ molecules are shown in Eqs.~(\ref{splitting-1})-(\ref{splitting-4}) in $\textrm{Appendix}$].
Furthermore, we use a model [see Appendix~C] to calculate the amplitudes of relevant NMR lines,
labelled by $\mathbf{A}_{\textrm{sim}}(\theta,\phi)=[\textrm{a}_1^{(\textrm{sim})}(\nu_{-1}), \textrm{a}_2^{(\textrm{sim})}(\nu_0),\textrm{a}_3^{(\textrm{sim})}(\nu_1)]$
when $\theta$ and $\phi$ are given.
Conversely, the parameters of $\theta$, $\phi$ can be obtained by minimizing
the norm $|\mathbf{A}_{\textrm{sim}}(\theta,\phi)-\mathbf{A}_{\textrm{exp}}|$.
However, multiple pairs of $(\theta, \phi)$ may result in the same spectrum if they satisfy certain symmetric conditions.
To remove this ambiguity,
we measure the nuclear-spin spectra and three sets of $\mathbf{A}_{\textrm{exp}}$
under various initial states $\rho_0$ by controlling the guiding field along ${x}$, ${y}$ and ${z}$~[\onlinecite{jiang2018experimental}], in turn.
The corresponding $\textrm{NMR}$ spectra are shown in Fig.~\ref{fig4}\textbf{b}.
The $\theta,\phi$ can be unambiguously determined by finding the simulated $\mathbf{A}_{\textrm{sim}}(\theta,\phi)$ that simultaneously best matches the measured $\mathbf{A}_{\textrm{exp}}$ under three initial states.
%In particular, the amplitude of zero-quantum line at $\nu_0$ markedly varies with initial spin states.
The magnetic field is determined to be such that $\theta \approx 1.289$~rad and $\phi \approx 0.047$~rad, and $B\approx 1.0788 \cdot 10^{-7}$~T.
The simulated Fourier spectra with red lines are in a good agreement with the experimental spectra.

%{\color{red}Here we note that the above multiparameter estimation currently requires three independent measurements to reconstruct the magnetic field. It is possible to prepare a linear combination of the initial spin states in previous three measurements by applying a guiding field along the $\frac{1}{\sqrt{3}}[\hat{x}+\hat{y}+\hat{z}]$ direction, and then perform a single-scan measurement of magnetic field vector.}
%This can significantly save measurement time.

In order to benchmark the precision of multiparameter estimation in our protocol,
we simulate the performance of $\theta$, $\phi$ deviation under the influence of spectral amplitude variance.
The variances $\{ \delta \textrm{a}_1^{(\textrm{exp})}, \delta \textrm{a}_2^{(\textrm{exp})}, $ $\delta \textrm{a}_3^{(\textrm{exp})}\}$
are assumed to obey Gaussian noise $\sim \mathcal{N}(0,\sigma ^2)$,
which comes mainly from the photon-shot noise of the probe beam in our experiment.
Figure~\ref{fig4}\textbf{c} shows the $\theta$, $\phi$ joint statistical distribution with $\sigma=0.01$ (corresponding to our current experimental signal-to-noise ratio (SNR) of 100).
The uncertainties (standard deviations) of $\theta$, $\phi$ values are respectively estimated to be $\sigma_{\theta}\approx0.009$ rad and $\sigma_{\phi}\approx0.017$~rad,
as shown in their histograms of deviations.
%Our simulation further shows that the uncertainties for determining parameters $\theta$, $\phi$ approximately scale with $\sigma_{\theta,\phi} \propto \sigma$
%when $\sigma$ ranges from 0 to 0.02.
%Thus, given a fixed measurement time ($\sim 3 \tau_{\textrm{coh}}$) to perform the sensor operation,
One could increase, for example, the nuclear-spin initial polarization and $\textrm{NMR}$ detector sensitivity, to decrease $\sigma$ and then achieve better orientation resolution.

We next discuss how precisely we can determine the magnitude of a magnetic field vector.
The magnetic magnitude can be achieved by exploiting the Zeeman splitting of strongly interacting nuclear spins,
for example,
using the splitting $\Delta$ in the formic acid spectrum.
For other $^{13}$CH$_n$ molecules, there exists similar Zeeman splitting that can be used to measure the magnetic field magnitude
[see Eqs.~(\ref{splitting-1})-(\ref{splitting-4}) in Appendix].
We use a set of such measurements to quantify the estimation precision.
%The measurement of a constant applied $2.9\cdot 10^{-8}~\textrm{T}$ field is repeated 300 times (30~s for each measurement).
The uncertainties of $\Delta$ in the frequencies extracted from spectral profile fits are in the order of $0.3~\textrm{mHz}$.
As shown in Fig.~\ref{fig4}\textbf{d}, the statistical standard deviation of absolute magnetic field is $10^{-11}~\textrm{T}$.
%Within the same molecule, Zeeman splittings of various subspaces could operate as a comagnetometer to enhance the precision~\cite{wu2018, Wu2019}.

%The above approach using formic acid can be directly applied to other molecules comprising of strongly interacting nuclear spins.
%Appropriate categories of sensing molecules are presented in Supplementary Information.

%\subsection{Gyroscope}

Nuclear spins can also be used to measure inertial rotation.
Conventional nuclear spin gyroscopes~\cite{donley2010nuclear, walker1997spin} usually rely on measuring the frequency shift of the freely precessing non-interacting spins (such as $^{3}$He) in a bias magnetic field,
limiting the gyroscope to detection of rotation along the magnetic field direction.
Here, we show that it could overcome this limitation by using strongly interacting nuclear spins, enabling to estimate three dimensional rotations.
Similar to magnetic field sensing, strongly interacting nuclear spins can also be sensitive to magnetic pseudo-fields originating from inertial rotation.
A magnetic pseudo-field $\mathbf{B}_{\Omega}^j=\bm{\Omega}/\gamma_j$ is induced on the $j$th spin~\cite{donley2010nuclear, wood2017magnetic,heims1962theory}
when the nuclear spin is probed with an atomic vapour sensor in a rotating frame, as shown in Fig.~\ref{rotation}\textbf{a}.
Here $\bm{\Omega}(\theta, \phi)$ is the rotation vector and parameterized with the polar angle $\theta$ and azimuthal angle $\phi$ and magnitude $\Omega$.
The corresponding Hamiltonian is
\begin{equation}
  \begin{cases}
    \mathcal{H} (\theta, \phi, \Omega)  = \mathcal{H}_{\Omega} + \mathcal{H}_{\textrm{spins}} + \mathcal{H}_{\textrm{int}},\\
  \mathcal{H}_{\Omega}=-  2\pi \bm{\Omega}(\theta,\phi) \cdot \sum_j  \mathbf{I}_j,
  \end{cases}
  \label{Hrotation}
  \end{equation}
where $\mathcal{H}_{\textrm{spins}}$ and $\mathcal{H}_{\textrm{int}}$ are described in Eq.~(\ref{Hamiltonian}).

Simulations are performed to investigate the rotational effect on the spectra of $^{13}$CH$_n$ molecules.
Figure~\ref{rotation}\textbf{b} shows the spectra of acetonitrile in different magnetic fields and a rotation along the $x$ axis.
With decreasing applied magnetic-field strength,
the resonance lines gradually change from a multiplet to a doublet.
More generally,
under inertial rotation and zero magnetic field, arbitrary molecules yield fewer resonance lines than those in real magnetic field
[see Eq.~(\ref{rotation-1}) in Appendix].
Thus, in contrast to real magnetic field sensing,
a variety of molecular species that have no overlapping lines can function as rotation sensors in zero field.

The spectra of strongly interacting nuclear spins can be used to the simultaneous estimation of the rotation speed and the orientation of the rotation axis.
As an illustration, we consider the case of nuclear spins initialized with the guiding field along $z$.
When the rotation axis is along the sensitive $z$-axis of atomic vapour sensor,
the resonance line of formic acid remains a single line (Fig.~\ref{rotation}\textbf{c}, bottom).
When the rotation axis is along $x$, the resonance line splits into a doublet with a splitting of $2\Omega$ (Fig.~\ref{rotation}\textbf{e}, bottom).
When the rotation axis is along $\theta=\pi/4$, $\phi=0$,
the resonance line splits into a triplet (Fig.~\ref{rotation}\textbf{d}, bottom).
%Under guiding fields along other axes, the rotationally induced spectra are presented in Fig.~\ref{rotation}\textbf{c}-\textbf{f} (top and middle).
The above described spectra can still be explained by the theories [see Eqs.~(\ref{splitting-1})-(\ref{splitting-4}) in Appendix] for conventional magnetic field,
if $\gamma_h \mathbf{B}$ and $\gamma_c \mathbf{B} $ are replaced by $ \bm{\Omega}$.
The measurement of the rotation vector $\bm{\Omega}$ can be performed with the same protocol as the one for the vector magnetometry above.
As demonstrations, Figure~\ref{rotation}\textbf{c}-\textbf{f} show the simulated spectra of formic acid for four physical rotations with different orientations of the rotation axis.
Clearly, they show significant differences.
Using the same procedure as for conventional field sensing,
we can obtain the rotation frequency and the orientation of the rotation axis.
Based on the experimental precision and orientation resolution of our sensor for the real magnetic field,
the precision of the rotation frequency is about $2.7\cdot 10^{-4}$~Hz (corresponding to the bias stability of gyroscope~\cite{donley2010nuclear}) and the rotation orientation resolution is about 0.02~rad.
With further optimization (see conclusion and outlook for details),
it should be possible to enhance the single-scan precision with several orders of magnitude improvement,
which is sufficient for the measurement of Earth's rotation (about $10^{-5}$~Hz).

\section{Conclusion and Outlook}

%We have demonstrated an interaction-based multiparameter metrology using the quantum resource of strong interactions among nuclear spins in molecules.
In this paper, we have introduced a new approach to multiparameter metrology, using a set of strongly interacting nuclear spins as a quantum resource.
Using this technique,
we have demonstrated a new type of sensor capable of detecting all three components of a magnetic field or the axis of rotation of an overall sample rotation.
This new measurement scheme has promising applications in
geophysical surveys, high-precision navigation, and testing fundamental physics.
Moreover, our technique naturally avoids the field contamination induced by previously-required reference fields~\cite{seltzer2004unshielded,patton2014all,thiele2018self,li2017dual},
and opens a feasible route towards realizing quantum noise-limited multiparameter sensors.

A further improvement of the experimental sensitivity can be anticipated.
For example, the probe states used in this work are thermal equilibrium states with low nuclear-spin polarization.
The extension to hyperpolarized states with near-unity spin polarization through many well-developed techniques, such as parahydrogen-induced polarization (PHIP)~\cite{adams2009reversible, theis2011parahydrogen}, dynamic nuclear polarization (DNP)~\cite{maly2008dynamic} and spin-exchange optical pumping (SEOP)~\cite{walker1997spin} techniques,
could yield at least five orders of magnitude improvements over currently achievable sensitivity.
Moreover, it is also possible to improve the readout sensitivity of the atomic vapour sensor.
For example, the application of magnetic gradiometer to the detection of nuclear-spin magnetization, as recently demonstrated in Ref.~\cite{jiang2019magnetic}, could enhance the multiparameter estimation sensitivity by a factor of about ten.
With recent quantum control technologies~\cite{jiang2018experimental,jiang2018numerical} in ultralow-field NMR~\cite{ledbetter2011near, jiang2020interference, tayler2017invited,jiang2021zero},
the NOON of the nuclear spins in each molecule can be created which offers an enhanced sensitivity to the multiparameter estimation
at a fundamental level~\cite{jones2009magnetic}.

Although demonstrated for strongly-interacting nuclear-spin systems,
the protocol presented here can be directly applied to other strongly interacting spin systems and design a new class of multiparameter quantum sensors.
%As we show in this work, we have successfully applied nuclear-nuclear spin clusters in ultralow field to realize vector magnetometer and gyroscope,
%and we show that such gyroscope can greatly improve the measurement accuracy.
We suggest future theoretical and experimental studies of the interaction-based multiparameter metrology with electron-nuclear~\cite{bermudez2011electron, zhao2011atomic, schweiger2001principles} or electron-electron~\cite{xiao2020magnetic, qin2016magnetic} interacting spin systems,
taking a fresh look at many techniques, such as spectroscopy, magnetometry, imaging, navigation, and even biocompass mechanism.
%such as alkali-metal atomic gas and nitrogen-vacancy centres~\cite{shi2014sensing,abobeih2019atomic}.
%The strong hyperfine-coupling strength between nuclear and electron spins can expand the magnetic field range where spin clusters are strongly coupled and sensitive to the field direction.
%These electron-nuclear spin clusters usually can achieve high polarization through optical pumping, and thus enable to measure magnetic field vector with high sensitivity.
For example, the application to nitrogen-vacancy centers~\cite{shi2014sensing,mamin2013nanoscale}, which offer the capability of realizing nanometer-scale spatial resolution, could realize a completely new nanometer-scale vector magnetic-field microscope.
It can also be applied to explore the magnetic-field orientation dependence of electron-electron spin systems,
such as radical pairs in magnetic proteins, which closely relate with the hitherto mysterious biocompass mechanism~\cite{xiao2020magnetic,qin2016magnetic}.
We are therefore convinced that this new type of interaction-based multiparameter quantum metrology opens a promising new field of research.

\section*{Acknowledgements}
We acknowledge Dmitry Budker, Ren-Bao Liu, \mbox{Roman P. Frutos} and Teng Wu for valuable discussions, and Jiankun Chen for drawing the set-up diagram.
This work was supported by National Key Research and Development Program of China (grant no. 2018YFA0306600), National Natural Science Foundation of China (grants nos. 11661161018, 11927811, 12004371), Anhui Initiative in Quantum Information Technologies (grant no. AHY050000),
USTC Research Funds of the Double First-Class Initiative (grant no. YD3540002002).

\section*{APPENDIX A: Experimental details}
\label{setup}
\subsection*{1. Experimental set-up}

As depicted schematically in Fig.~\ref{fig1}\textbf{a},
%The spatial axes of the experimental set-up are represented as $x^{'}y^{'}z^{'}$.
the liquid ($\sim 100$ $\mu$L) contained in a $5$-mm $\textrm{NMR}$ tube is polarized in a Halbach magnet ($B_p \approx1.3~\textrm{T}$),
and then pneumatically shuttled to a sensing and readout region ($\sim 1$~mm above the atomic vapour cell) inside a five-layer $\mu$-metal magnetic shield.
During the pneumatic transfer,
a guiding solenoid and an additional set of mutually orthogonal Helmholtz coils are used
to generate a guiding field along an axis (${x}$, ${y}$ or ${z}$) for preparing the initial spin state $\rho_0$, as described in Eq.~(\ref{rho0}).
The magnitude of the guiding field ($B_g\approx 10^{-4}~\textrm{T}$) satisfies $B_g \gg J /(\gamma_h-\gamma_c)$.
After the shuttling,
the guiding field is switched off within 10~$\mu$s.
The nuclear spins of the liquid then evolves under the external magnetic field and, accordingly,
generate an oscillating magnetization signal detected with a $^{87}$Rb vapour sensor~\cite{tayler2017invited, ledbetter2011near, jiang2018experimental, jiang2019magnetic}.
The $^{87}$Rb vapour is resistively heated to $180$ $^\circ$C.
As shown in Fig.~\ref{fig1}\textbf{c},
the $^{87}$Rb atoms in a vapour are pumped with a circularly polarized laser beam propagating in the $x$ direction.
The laser frequency is tuned to the center of the buffer-gas (N$_2$) broadened and shifted $\textrm{D}1$ line.
The magnetic field is measured via optical rotation of linearly polarized probe laser light at the $\textrm{D}2$ transition propagating in the $y$ direction.
The sensitivity of the $^{87}$Rb vapour sensor along the $z$ axis is optimized to $\approx 20~\textrm{fT} \cdot \textrm{Hz}^{-1/2}$.

\subsection*{2. Strongly-interacting nuclear-spin samples}
%Our $\textrm{NMR}$ sensor protocol commonly exploits the strongly coupled $^{13}$CH$_n$ clusters,
%where a $^1$H spin couples to a $^{13}$C spin with the same scalar spin-spin coupling strength $J$ in the ultralow-field regime.
As used in our experiments,
the liquids including formic acid (H$^{13}$COOH), formaldehyde ($^{13}$CH$_2$O), acetonitrile ($^{13}$CH$_3$CN), and acetic acid ($^{13}$CH$_3$COOH)
are from Sigma-Aldrich, and are contained in standard $5$-mm $\textrm{NMR}$ tubes.
%The $\textrm{NMR}$ tubes are flame-sealed under vacuum following five freeze-pump-thaw cycles
%in order to remove dissolved oxygen, which causes additional spin relaxation.
The samples were flame-sealed under vacuum in
NMR tubes after five freeze-pump-thaw cycles to remove dissolved
oxygen, which is otherwise a significant source of relaxation.
Nuclear spin coherence times $\tau_{\textrm{coh}}\approx 10.4$~s and scalar spin-spin coupling strengths $J\approx 222.2$~Hz for formic acid,
$\tau_{\textrm{coh}}\approx 0.8$~s and $J\approx 163.9$~Hz for formaldehyde, $\tau_{\textrm{coh}}\approx 4.7$~s and $J\approx 136.25$~Hz for acetonitrile,
and $\tau_{\textrm{coh}}\approx 8.8$~s and $J\approx 129.5$~Hz for acetic acid.

\section*{APPENDIX B: Splitting of strongly interacting spins}

\subsection*{1. Zeeman splitting}

When the magnetic field is sufficiently smaller than the scalar spin-spin couplings (i.e., $\gamma_h B, \gamma_c B $ $ \ll |J|$),
the nuclear spins in a $^{13}$CH$_n$ molecule are strongly coupled to each other.
In this situation, the good quantum numbers are the total spin angular momentum $f=f_c+f_h$
and the corresponding magnetic quantum number $m_f$ with possible values $-f\leq m_f \leq f$.
Here,
$f_c=\pm 1/2$ and $f_h=n/2-k$ are respectively the total $^{13}$C and H spin angular momentum,
and $k=0,1,...,n/2$ for even $n$ or $k=0,1,...,(n-1)/2$ for odd $n$.
The eigenstates of $^{13}$CH$_n$ molecules are denoted by $|{fm_f;k}\rangle$~[\onlinecite{ledbetter2011near}, \onlinecite{appelt2010paths}],
which can be approximately those eigenstates of the zero-field Hamiltonian $\mathcal{H}=\sum_{j>i}^n {{ J_{ij}}}  \mathbf{I}_i \cdot \mathbf{I}_j$.
The examples of $^{13}$CH, $^{13}$CH$_2$, $^{13}$CH$_3$ energy levels are presented in Fig.~\ref{fig2}.

%The knowledge of the Zeeman splitting of sensor molecules is curial for determining the magnitude of magnetic field or rotation frequency.
First, we consider the case of real magnetic fields.
When the ensemble of $^{13}$CH$_n$ molecules are measured parallel to the external magnetic field,
the selection rules for allowed transitions are $\Delta f=0,\pm 1$, $ \Delta m_f=0$.
Because the protons are equivalent in a $^{13}$CH$_n$,
there is an additional selection rule $\Delta k=0$~[\onlinecite{ledbetter2011near}].
Accordingly,
one could observe $n-2k$ resonance lines centered at $\frac{1}{2}J(1+n-2k)$, with the frequencies
\begin{equation}
\begin{array}{l}
\nu_{f,m_f;k}^{f',m_f;k}=\frac{1}{2}J(1+n-2k)+\frac{2m_f (-\gamma_h+\gamma_c)}{1+n-2k}B,
\end{array}
\label{splitting-1}
\end{equation}
where $\nu_{f,m_f;k}^{f',m'_f;k'}$ denotes the transition frequency between the eigenstates $|{fm_f;k}\rangle$ and $|{f'm'_f;k'}\rangle$.
The adjacent zero-quantum Zeeman splitting is
\begin{equation}
\begin{array}{l}
\Delta_{\textrm{ZQ}}=\frac{2}{1+n-2k}(\gamma_h-\gamma_c)B.
\end{array}
\label{splitting-2}
\end{equation}

When the ensemble of $^{13}$CH$_n$ molecules are measured perpendicular to the external magnetic field,
the selection rules for allowed transitions are $\Delta f=0,\pm 1$, $ \Delta m_f=\pm 1$, and $\Delta k=0$.
Accordingly, one could observe $2(n-2k)$ resonance lines centered at $\frac{1}{2}J(1+n-2k)$, with the frequencies
\begin{equation}
\begin{array}{l}
\nu_{f,m_f;k}^{f',m_f\pm1;k}=\frac{1}{2}J(1+n-2k)+[\frac{2m_f (-\gamma_h+\gamma_c)}{1+n-2k} \pm \frac{(n-2k)\gamma_h+\gamma_c}{1+n-2k}]B.
\end{array}
\label{splitting-3}
\end{equation}
%There are $2(n-2k)$ spectral lines centered at $\frac{1}{2}J(1+n-2k)$.
The Zeeman splitting of the central doublet, which have the highest signal-to-noise ratio compared to others, is
\begin{equation}
\begin{array}{l}
\Delta_{\textrm{SQ}}=\frac{2}{1+n-2k}[\gamma_h+(n-2k)\gamma_c]B.
\end{array}
\label{splitting-4}
\end{equation}
For formic acid with $n=1$ and $k=0$, the Zeeman splitting of the doublet is $\Delta=(\gamma_c+\gamma_h)B$,
as used in the text to determine the strength of magnetic field.
The above discussions only consider the first-order Zeeman effects of $^{13}$CH$_n$ molecules.
The higher-order Zeeman effects cause a systematic error of measuring magnetic-field strengths and can be seen in Refs.~\cite{appelt2010paths,wu2018nuclear}.

\subsection*{2. Rotation-induced splitting}

We now consider the splitting of $^{13}$CH$_n$ molecules under an inertial rotation.
The magnetic pseudo-field induced by the inertial rotation is described in the text.
In the absence of real magnetic field,
the above Eqs.~(\ref{splitting-1})-(\ref{splitting-4}) can also be used in the case of magnetic pseudo-fields originating from inertial rotation.
In this case, $\gamma_h B$ and $\gamma_c B $ should be replaced by $ \Omega$.
Remarkably,
in contrast to the case of real magnetic field,
$\nu_{f,m_f;k}^{f',m_f;k}$ and $\nu_{f,m_f;k}^{f',m_f\pm1;k}$ become independent of the magnetic quantum number $m_f$.
As a result, the spectra of molecules under inertial rotation and zero field are very simple:
\begin{equation}
\begin{cases}
\nu_{f,m_f;k}^{f',m_f;k}(\Omega)=\frac{1}{2}J(1+n-2k),\\
\Delta_{\textrm{ZQ}} (\Omega)=0,\\
\nu_{f,m_f;k}^{f',m_f\pm1;k} (\Omega) =\frac{1}{2}J(1+n-2k) \pm \Omega,\\
\Delta_{\textrm{SQ}} (\Omega) =2\Omega.
\end{cases}
\label{rotation-1}
\end{equation}
(1) ``$\Delta_{\textrm{ZQ}}=0$'' implies that zero-quantum transition lines remain single lines under rotation,
whereas zero-quantum lines are usually split into a multiplet in conventional magnetic fields,
such as those shown in Fig.~\ref{fig2}\textbf{c},\textbf{e}.
(2) ``$\Delta_{\textrm{SQ}}=2\Omega$'' implies that single-quantum transition lines are doublets with a splitting of $2\Omega$.
This is significantly different from that in real magnetic fields,
where the spitting depends on $n$ and $k$ and the number of splitting lines is usually large (Fig.~\ref{fig2}\textbf{b},\textbf{d},\textbf{f}).
%A full simulation of $^{13}$CH$_n$ molecules under inertial rotation can be found in Supplementary Information.

\section*{APPENDIX C: Multiparameter estimation process}

The multiparameter estimation presented in this work includes three stages: probe preparation, encoding with the strongly interacting spins, and probe readout.
In the following, we provide the detailed theory of such three stages.

\subsection*{1. Probe preparation}

As described in the part of experimental set-up in Appendix~A1,
the nuclear spins are polarized in a permanent magnet and then pneumatically shuttled to a sensing region.
When the guiding field is turned off within 10~$\mu$s,
the spin state remains the high-field equilibrium state.
In the high-temperature approximation,
the relevant initial spin state is~\cite{jiang2018experimental}
\begin{equation}
\begin{array}{l}
\rho_0=\mathbbm{1}/2^n-\sum_j \varepsilon_j \mathbf{I}_j \cdot \hat{\mathbf{k}}_g,
\end{array}
\label{rho0}
\end{equation}
where $\varepsilon_j=\gamma_jB_p/k_\textrm{B} T \sim 10^{-6}$, $k_\textrm{B}$ is the Boltzmann constant,
$T$ is the temperature,
$\mathbf{I}_j = (\hat{I}_{jx}, \hat{I}_{jy}, \hat{I}_{jz})$ is the spin angular momentum operator of the $i$th spin,
and $\hat{\mathbf{k}}_g$ is the unit vector along the guiding field orientation.
For example, $\hat{\mathbf{k}}_g=(0,0,1)$ corresponds to the case of $z$-guiding field.
Thus, the initial state can be controlled by the guiding field direction $\hat{\mathbf{k}}_g$.
%which can be simply controlled by the guiding field coils.

\subsection*{2. Encoding unknown parameters}

Here we use the encoding of an unknown magnetic field as an example to describe the basic encoding process.
The process is general and suitable for encoding inertial rotation.
In Appendix B, we have described the process to determine the strength of an unknown magnetic field or inertial rotation.
Thus, we focus on how to estimate their orientation in this part.
The orientation of magnetic field $\mathbf{B}$ is parameterized with the polar angle $\theta$ and azimuthal angle $\phi$,
as shown in the inset of Fig.~\ref{fig1}\textbf{a}.
In the measured magnetic field,
subsequent evolution of the quantum state of the spins in each molecule produces oscillating magnetization,
which is detected by the $^{87}$Rb vapour sensor.
The observable in our experiment is the total $z$ magnetization,
\begin{equation}
S(\rho_0, \theta, \phi, t)\propto \textrm{Tr} [\rho(\theta, \phi, B, t) \sum_j \gamma_j \hat{I}_{jz}],
\end{equation}
where $\rho(\theta, \phi, B, t)$ is the time dependent density matrix, as described in Eq.~(\ref{erho}).
Consequently,
the component of the sensor signal, evolving with the frequency of $\nu_{f,m_f;k}^{f',m_f';k}$, can be written with
\begin{equation}
\begin{array}{l}
S_{fm_f;k}^ {f'm_f';k'} (\rho_0,\theta,\phi, t) \propto \\
   \Re_{fm_f;k}^ {f'm_f';k'}(\rho_0,\theta,\phi) \cdot \textrm{cos}(\Phi-2\pi \nu_{f,m_f;k}^{f',m_f';k'} t)e^{-t/\tau_{\textrm{coh}}},
\end{array}
\label{signal}
\end{equation}
where the amplitude $\Re_{fm_f;k}^ {f'm_f';k'}(\rho_0,\theta,\phi)$ and phase $\Phi$ are defined by the following relation
\begin{equation}
\begin{array}{l}
\Re_{fm_f;k}^ {f'm_f';k'}(\rho_0,\theta,\phi) e^{i{\Phi}}\equiv
\\
\langle {fm_f;k}|\rho_0|{f'm_f';k'}\rangle \langle {f'm_f';k'}| \hat{\mathbf{O}}_z|{fm_f;k}\rangle,
\end{array}
\label{amp}
\end{equation}
the spin operator $\hat{\mathbf{O}}_l \equiv \sum_j \gamma_j \hat{I}_{jl}$  ($l=x,y,z$),
and $e^{-t/\tau_{\textrm{coh}}}$ is due to the decoherence effect.
The amplitude $\Re_{fm_f;k}^ {f'm_f';k'}(\rho_0,\theta,\phi)$  can be determined by the observed NMR spectrum, as seen in ``Probe readout''.

%We note that $\Re_{fm_f;k}^ {f'm_f';k'}(\rho_0,\theta,\phi)$ corresponds to the amplitude of the resonance line at $\nu_{f,m_f;k}^{f',m'_f;k'}$

To achieve the clear relation between the amplitude $\Re_{fm_f;k}^ {f'm_f';k'}(\rho_0,\theta,\phi)$ and the unknown parameters $\theta$, $\phi$,
we introduce a convenient orthonormal coordinate system,
defined by using only $\theta$ and $\phi$.
In such a coordinate system (see the inset of Fig.~\ref{fig1}\textbf{a}), the quantization axis ($\mathbf{z}'$) is chosen along the magnetic field, i.e.,
\begin{equation}
\begin{cases}
\mathbf{z'}\equiv\mathbf{B}/B=(\sin \theta \cos \phi, \sin\theta \sin \phi, \cos \theta),\\
\mathbf{x'}\equiv\frac{\partial \mathbf{z'}}{\partial \theta}=(\cos \theta \cos \phi, \cos\theta \sin\phi, -\sin\theta),\\
\mathbf{y'}\equiv(\frac{1}{\sin \theta})\frac{\partial \mathbf{z'}}{\partial \phi}=(-\sin \phi, \cos\phi,0).
\end{cases}
\label{coordinate-1}
\end{equation}

%As a consequence of the coordinate transformation,
%the spin operator $\hat{\mathbf{O}}_l$ can be expressed by the spin operator $\hat{\mathbf{O}}'_{l'}=\sum_j \gamma_j \hat{I}_{j,l'}$ ($l'=x',y',z'$),
Thus the spin operator in the laboratory frame can be transformed to $\hat{\mathbf{O}}'_{l'}=\sum_j \gamma_j \hat{I}_{jl'}$ ($l'=x',y',z'$) by
%which is defined the new defined coordinate system.
\begin{equation}
\hat{\mathbf{O}} =\mathcal{P}(\theta,\phi)  \hat{\mathbf{O}}',
\label{transf}
\end{equation}
where $\hat{\mathbf{O}}=(\hat{\mathbf{O}}_x, \hat{\mathbf{O}}_y, \hat{\mathbf{O}}_z)^{T}$,
$\hat{\mathbf{O}}'=(\hat{\mathbf{O}}'_{x'}, \hat{\mathbf{O}}'_{y'}, \hat{\mathbf{O}}'_{z'})^{T}$,
and $\mathcal{P}(\theta,\phi)$ is a $3\times3$ transformation matrix
\begin{equation}
\mathcal{P}(\theta,\phi)=\begin{pmatrix}
\cos\theta \cos\phi & -\sin \phi & \sin \theta \cos \phi\\
\cos\theta \sin\phi & \cos \phi & \sin \theta \sin \phi\\
-\sin\theta  & 0 & \cos \theta
\end{pmatrix}.
\label{P}
\end{equation}
Further, the element $\langle {f'm_f';k}| \hat{\mathbf{O}}_l|{fm_f;k}\rangle$ in the laboratory frame
can be expressed in the new defined coordinate system as
\begin{equation}
\langle {f'm_f';k}| \hat{\mathbf{O}}|{fm_f;k}\rangle=\mathcal{P}(\theta,\phi) \cdot \langle {f'm_f';k}| \hat{\mathbf{O}}'|{fm_f;k}\rangle.
\label{transf}
\end{equation}

Moreover, the $\langle {f'm_f';k}| \hat{\mathbf{O}}'|{fm_f;k}\rangle$ is independent of the orientation of the magnetic field ($\theta,\phi$)
and thus can be calculated without knowing any information of the magnetic field $\mathbf{B}$ in ultralow-field regime.
The detailed methods to calculate $\langle {f'm_f';k}| \hat{\mathbf{O}}'|{fm_f;k}\rangle$ can be seen in Ref.~[\onlinecite{sjolander2016transition}].
For the typical $^{13}$CH molecules, for example, we have
\begin{equation}
\begin{cases}
\langle {0,0;0}| \hat{\mathbf{O}}'|{1,0;0}\rangle=(0,0,\frac{\gamma_c-\gamma_h}{2})^{T},\\
\langle {0,0;0}| \hat{\mathbf{O}}'|{1,-1;0}\rangle=( \frac{\gamma_c-\gamma_h}{2\sqrt{2}} , -\frac{i(\gamma_c-\gamma_h)}{2\sqrt{2}} ,0)^{T},\\
\langle {0,0;0}| \hat{\mathbf{O}}'|{1,1;0}\rangle=( \frac{-\gamma_c+\gamma_h}{2\sqrt{2}} , -\frac{i(\gamma_c-\gamma_h)}{2\sqrt{2}} ,0)^{T}.\\
\end{cases}
\label{CH}
\end{equation}

Based on Eqs.~(\ref{transf}) and (\ref{CH}),
$\langle {f'm_f';k}| \hat{\mathbf{O}}_z|{fm_f;k}\rangle$ can be represented by the only parameters $\theta$ and $\phi$, i.e.,
\begin{equation}
\begin{array}{l}
\langle {f'm_f';k}| \hat{\mathbf{O}}_z|{fm_f;k}\rangle=
\hat{\mathbf{z}} \cdot  \mathcal{P}(\theta,\phi) \cdot \langle {f'm_f';k}| \hat{\mathbf{O}}'|{fm_f;k}\rangle,
\label{OT}
\end{array}
\end{equation}
where the unit vector $\hat{\mathbf{z}}=(0,0,1)$.
For the density matrix $\rho_0$, the element $ \langle {fm_f;k}|\rho_0|{f'm_f';k'}\rangle$ has similar result
\begin{equation}
\begin{array}{l}
\langle {f'm_f';k}| \rho_0 |{fm_f;k}\rangle= \\
(-B_p/k_\textrm{B} T)  \hat{\mathbf{k}}_g \cdot \mathcal{P}(\theta,\phi) \cdot \langle {f'm_f';k}| \hat{\mathbf{O}}'|{fm_f;k}\rangle.
\end{array}
\label{rhoT}
\end{equation}

According to Eqs.~(\ref{amp}), (\ref{CH}), (\ref{OT}), (\ref{rhoT}),
the amplitude $\Re_{fm_f;k}^ {f'm_f';k'}(\rho_0,\theta,\phi)$ can be finally represented by the two unknown parameters $\theta$ and $\phi$, i.e.,
\begin{equation}
\begin{array}{l}
\Re_{fm_f;k}^ {f'm_f';k'}(\rho_0,\theta,\phi) = \\
B_p/k_\textrm{B} T \cdot
|\hat{\mathbf{k}}_g \cdot \mathcal{P}(\theta,\phi) \langle {f'm_f';k}| \hat{\mathbf{O}}'|{fm_f;k}\rangle |\times  \\ |\hat{\mathbf{z}} \cdot  \mathcal{P}(\theta,\phi) \langle {f'm_f';k}| \hat{\mathbf{O}}'|{fm_f;k}\rangle|.
\end{array}
\label{R}
\end{equation}

Equation~(\ref{R}) is the central result of our interaction-based multiparameter estimation.
We make two important observations.
(1) $\Re_{fm_f;k}^ {f'm_f';k'}(\rho_0,\theta,\phi)$ is represented with two parts, $\theta$, $\phi$-dependent [i.e., $\mathcal{P}(\theta,\phi)$] and $\theta$, $\phi$-independent one [e.g., $\langle {f'm_f';k}| \hat{\mathbf{O}}'|{fm_f;k}\rangle$].
This clearly shows the relation between the multiple unknown parameters and $\Re_{fm_f;k}^ {f'm_f';k'}(\rho_0,\theta,\phi)$ that can be experimentally determined in NMR spectrum; see ``Probe readout'' for details.
(2) The use of strong interactions between nuclear spins permits nonzero transition probability $\langle {f'm_f';k}| \hat{\mathbf{O}}'_{z'}|{fm_f;k}\rangle$ (corresponding to zero-quantum transition; see Fig.~\ref{fig2}\textbf{a},\textbf{c}, \textbf{e}), offering an efficient route to encode the unknown parameters.

\subsection*{3. Probe readout}

The amplitude $\Re_{fm_f;k}^ {f'm_f';k'}(\rho_0,\theta,\phi)$ can be experimentally determined with the observed NMR spectrum, detected with a $^{87}$Rb vapour sensor.
One detailed example with formic acid molecules is presented in Fig.~\ref{fig4}b.
Specifically, the amplitudes of zero- and single-quantum lines for $^{13}$CH molecules,
$\Re_{0,0;0}^ {1,0;0}(\rho_0,\theta,\phi)$ and $\Re_{0,0;0}^ {1, \pm 1;0}(\rho_0,\theta,\phi)$
can be determined from the triplet (see Fig.~\ref{fig4}b),
and lastly can be used to extract the parameters $\theta$ and $\phi$ of the magnetic field $\mathbf{B}(\theta,\phi)$.

%It is essential to measure the above-discussed $\Re_{fm_f;k}^ {f'm_f';k'}(\rho_0,\theta,\phi)$ that encodes the parameters $\theta$ and $\phi$ to be measured.

%\end{methods}

%\noindent
%\textbf{Author contributions}
%M.J. designed the experimental protocols, performed numerical simulations, analyzed the data and wrote the manuscript.
%Y.L.J., Q.L. performed experiments and analyzed the data.
%D.S. proofread and edited the manuscript.
%X.H.P. provided the overall management of the project, contributed to the design of the experiment and wrote the manuscript.
%All authors contributed with discussions and to the final form of the manuscript.

%\noindent
%\textbf{Competing Interests}
%The authors declare that they have no competing financial interests.

\bibliographystyle{naturemag}
\bibliography{mainrefs}

\end{document}